\documentclass{article}

\usepackage{arxiv}

\usepackage{url}            
\usepackage[ruled,noline]{algorithm2e}
\usepackage{amsmath}
\usepackage{amsfonts}
\usepackage[colorinlistoftodos]{todonotes}
\usepackage{lineno}
\usepackage{float}
\usepackage{url}            

\usepackage{etoolbox}
\providetoggle{fullBackground}

\usepackage{parskip}  

\newcommand{\subjto}{\text{subject to}}
\newcommand{\prox}{\text{prox}}

\DeclareFontFamily{U}{wncy}{}
\DeclareFontShape{U}{wncy}{m}{n}{<->wncyr10}{}
\DeclareSymbolFont{mcy}{U}{wncy}{m}{n}
\DeclareMathSymbol{\comb}{\mathord}{mcy}{"58} 

\usepackage{accents}
\newlength{\dhatheight}

\title{Di-chromatic Interpolation of Magnetic Resonance Metabolic Imagery}

\author{
  Nicholas Dwork\thanks{www.nicholasdwork.com, nicholas.dwork@ucsf.edu} \\
  Department of Radiology and Biomedical Imaging \\
  University of California in San Francisco
    \And
  Jeremy W. Gordon \\
  Department of Radiology and Biomedical Imaging \\
  University of California in San Francisco
    \And
  Shuyu Tang \\
  Department of Radiology and Biomedical Imaging \\
  University of California in San Francisco
    \And
  Daniel O'Connor \\
  Department of Mathematics and Statistics \\
  University of San Francisco
    \And
  Esben Søvsø Szocska Hansen \\
  Department of Clinical Medicine \\
  Aarhus University
    \And
  Christoffer Laustsen \\
  Department of Clinical Medicine \\
  Aarhus University
    \And
  Peder E. Z. Larson \\
  Department of Radiology and Biomedical Imaging \\
  University of California in San Francisco
}

\begin{document}
\maketitle

\begin{abstract}
  Magnetic resonance imaging with hyperpolarized contrast agents can provide unprecedented \textit{in-vivo} measurements of metabolism, but yields images that are lower resolution than that achieved with proton anatomical imaging.
  In order to spatially localize the metabolic activity, the metabolic image must be interpolated to the size of the proton image.
  The most common methods for choosing the unknown values rely exclusively on values of the original un-interpolated image.  In this work, we present an alternative method that uses the higher-resolution proton image to provide additional spatial structure.  The interpolated image is the result of a convex optimization algorithm which is solved with the Fast Iterative Shrinkage Threshold Algorithm (FISTA).  Results are shown with images of hyperpolarized pyruvate, lactate, and bicarbonate using data of the heart and brain from healthy human volunteers, a healthy porcine heart, and a human with prostate cancer.
\end{abstract}

\keywords{interpolation \and image processing \and MRI \and spectroscopy}

\textbf{This work has been published in \textit{Magnetic Resonance Materials in Physics, Biology and Medicine}.}

\section{Introduction}
\label{sec:intro}


Magnetic Resonance Imaging (MRI) following injection of hyperpolarized compounds has permitted the investigation of metabolism in a non-invasive way without ionizing radiation \cite{ardenkjaer2003increase,golman2006real,schroeder2011hyperpolarized}.
Imaging of hyperpolarized Carbon-13 ($^{13}$C) has been shown to be valuable for cancer staging and treatment evaluation purposes \cite{nelson2013metabolic,kurhanewicz2019hyperpolarized,grist2019quantifying}, and cardiac evaluation \cite{cunningham2016hyperpolarized}.
As the carbon atoms move between compounds via metabolism, the relative amounts of each metabolite are observed based on their spectral chemical shift.
For example, with a hyperpolarized [1-$^{13}$C] pyruvate imaging study, one can image pyruvate, lactate, and bicarbonate to assess the cellular metabolism of carbohydrates \cite{larson2018investigation,golman2008cardiac}. 

Unlike conventional MRI of protons, where the signal returns to an equilibrium state that can be repeatedly measured, the hyperpolarized nuclei have a finite amount of longitudinal magnetization that can be used for imaging.
Due to the limiting signal decay rate of the hyperpolarized nuclei and the low signal-to-noise ratio of the metabolic byproducts, the resolution of the metabolic images is typically lower by approximately a factor of $5$ to $10$.

A metabolic imaging study typically consists of at least two sets of images: a high resolution proton image (necessarily weighted by proton density and possibly additionally weighted by other physical characteristics, e.g. $T_1$ and/or $T_2$) and a set of low resolution images, one for each metabolite containing the hyperpolarized atom.
Interpolation is used to enlarge the metabolic images to the size of the proton image which permits the localization of the metabolic activity.  Additionally, the enlarged metabolic image is often made into a false color image and fused with the proton image for improved localization \cite{wang2019hyperpolarized}.  The most common interpolation algorithms used include nearest-neighbor, linear, and sinc (which can be accomplished by zero-filling in the Fourier domain).
These methods rely exclusively on the values of the un-interpolated metabolic image and do not take advantage of the high-resolution proton image.  In this manuscript, we present an algorithm that does.


Other problems where an image of one modality is used to improve the resolution of another include MR and positron emission tomography (PET) \cite{knoll2016joint}, computed tomography (CT) and PET \cite{cui2018wavelet}, and MR and CT \cite{ehman2017pet}.
Existing techniques aim to simultaneously reconstruct both images from the raw data: the sinogram of CT, the Fourier samples of MR, and the projection data of PET.  We accept, as input, magnitude images while prior work accepts raw complex data.  The problems differ fundamentally: prior work is attempting to reconstruct both images from raw data with high fidelity while we are trying to interpolate one image using the information of another.

Interpolating directly from the magnitude images has several advantages.  Doing so eliminates the need to model the physics of the MRI system.  For a gradient based optimization solution, this eliminates two multiplications by sensitivity maps, and a Fourier transform and an inverse Fourier transform for each coil in each iteration of the optimization algorithm \cite{fessler2010model}.  By avoiding these operations, the proposed method is computationally efficient.  The resulting image in all cases is a magnitude image; thus, it is sensible to regularize the interpolated magnitude image. If one were to include data consistency of the raw complex data in the objective but regularize the interpolated magnitude image, the resulting optimization problem would be non-convex which could yield a sub-optimal solution.  By working directly with the magnitude images, we are able to regularize accordingly while presenting a convex optimization problem which maintains theoretical guarantees of optimality \cite{boyd2004convex}.  Finally, by working with reconstructed images, the algorithm can be applied to cases where the data has already been collected but the raw data was not preserved.

In this work, we present a new interpolation scheme for the metabolic images where the higher resolution image is used to inform the interpolation values.
By doing so, the spatial localization of the metabolic activity is made more apparent.
The interpolated image is the solution of a constrained convex optimization algorithm, which is efficiently solved with the Fast Iterative Shrinkage Threshold Algorithm (FISTA).  Matlab routines for this project are available at the author’s website: www.nicholasdwork.com.

\section{Methods}
\label{sec:methods}
\begin{figure}[ht]
  \begin{center}
    \includegraphics[width=0.5\linewidth]{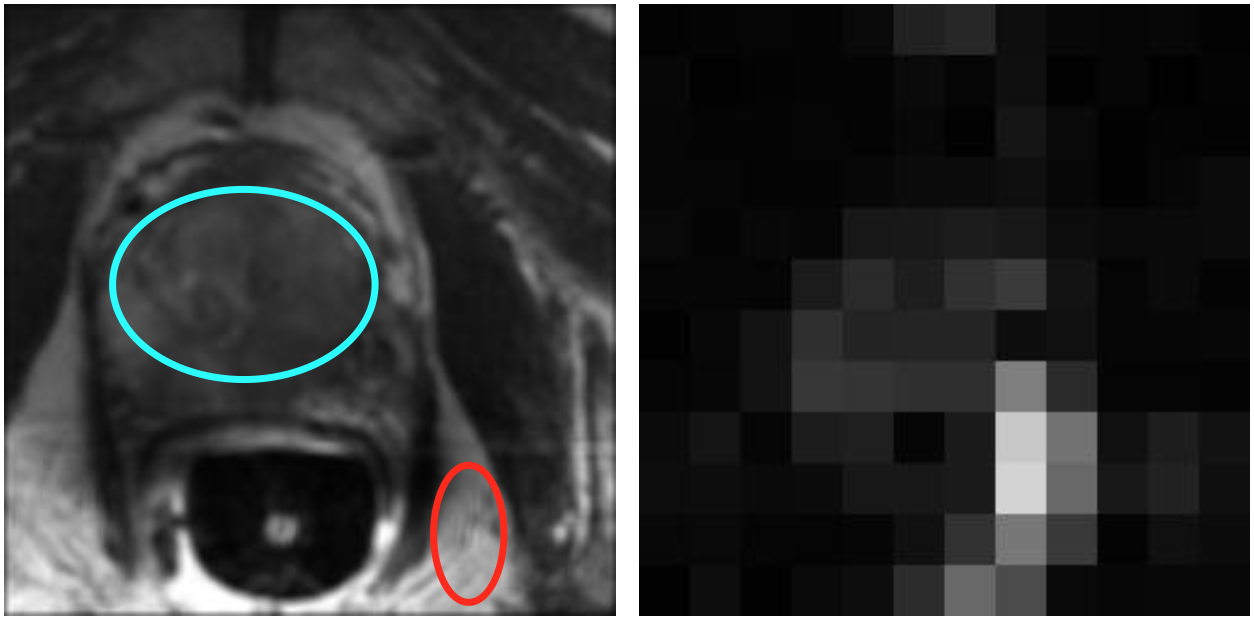}
  \end{center}
  \caption{ \label{fig:sampleData}  A representative pair of MR (left) $T_2$ weighted proton and (right) hyperpolarized Pyruvate images from a patient with prostate cancer. Note the significantly high resolution and resulting details in the proton image.  The cyan oval encircles a region of the prostate and the red oval encircles a region of fat. }
\end{figure}

Consider the sample data shown in Fig. \ref{fig:sampleData}; the high resolution anatomical image is shown on the left and the low resolution metabolic image is shown on the right.
The high resolution anatomical image is necessarily weighted by proton density; therefore, there is only signal where there is tissue.  However, there may be additional contrast imposed on the image as well (e.g. $T_2$ or $T_1$ contrast); we will address this phenomenon in section \ref{sec:intro:addtlContrast}.
We desire an image of the resolution of the proton image that presents the intensity of the metabolic image.  We know that if we average and downsample the interpolated image, the resulting intensity should be close to the intensities of the low resolution image.  That is, we will assume the following degradation model (commonly used with optical interpolation in pan-sharpening algorithms \cite{li2009fusion,garzelli2016review,wu2019remote}):
\begin{equation}
  M = D \, B \, I_M + n,
  \label{eq:degModel}
\end{equation}
where $M\in\mathbb{R}^{\mathcal{M}_M\times \mathcal{N}_M}$ is the low resolution metabolic image, $I_M$ is the interpolated metabolic image, $B$ is a circulant matrix that represents convolution with the blur kernel, $D$ represents the downsampling operator, and $n$ is additive Gaussian noise.
This is an underdetermined linear system.

If we imagine that each square block of color in the low resolution image were a square piece of putty or dough and we were asked to mold it so that it would look like the anatomy, intuitively, we would shape it to have the same contours, hills, and valleys as the high resolution proton image.  This intuition suggests that we should incorporate the gradient vectors of the anatomical image into the interpolation.  While constraining the solution with the physical equation of \eqref{eq:degModel}, we will encourage the gradients of the interpolated metabolic image to be similar to those of the high resolution anatomical image.  This idea is commonly used with bayer-patterned demosaicing in optical imaging \cite{lukin2004high}.

In order to make the gradient vectors meaningful for the low-resolution image, though, we will first have to accommodate differences in the dynamic ranges of the images, which may be different due to different coil geometries, excitation, or receiver gain.
Both images are scaled so that their values lie in the $[0,1]$ interval: $A=\hat{A}/\max(\hat{A})$ and $M=\hat{M}/\max(\hat{M})$, where $\hat{A}\in\mathbb{R}^{\mathcal{M}_A\times \mathcal{N}_A}$ and $\hat{M}\in\mathbb{R}^{\mathcal{M}_M\times \mathcal{N}_M}$ are the original anatomical and metabolic images, respectively.  The values $\mathcal{M}_M$ and $\mathcal{N}_M$ are the number of rows and columns of the metabolic image, respectively, and $\times$ represents the Cartesian cross product.  Similarly, $\mathcal{M}_A$ and $\mathcal{N}_A$ are the number of rows and columns in the anatomical image.
The gradient vectors of the scaled images can be used to inform the interpolation.

\subsection{Constructing the optimization problem}
We employ two separate methods for controlling how much the gradient vectors of the anatomical image and the interpolated image should match: a global user specified parameter $\lambda$, and a pixel-based weighting.
To determine the interpolated image, one solves the following convex optimization problem.
\begin{equation}
  \begin{aligned}
    \underset{I_M}{\text{minimize}} &\hspace{8pt} \frac{1}{2\,N_L} \| D\,B \, I_M - M \|_{Fr}^2 +
      \frac{\lambda}{2\,N_H}\left\| \nabla I_M - \nabla A \right\|_{w,2}^2 \\
    \subjto &\hspace{8pt} 0 \leq I_M \leq 1
  \end{aligned}
  \label{eq:diChromProb}
\end{equation}
where $\lambda>0$ is a regularization parameter, $A\in\mathbb{R}^{\mathcal{M}_A\times \mathcal{N}_A}$ is the high resolution anatomical image, and $\|\cdot\|_{Fr}$ represents the Frobenius norm (the square root of the sum of the matrix's elements squared).
The scalars $N_A=\mathcal{M}_A\,\mathcal{N}_A$ and $N_M=\mathcal{M}_M\,\mathcal{N}_M$ represent the number of pixels in $A$ and $M$, respectively.  The $(1/(2\,N_M)) \| D \, B \, I_M - M\|_{Fr}^2$ term in the objective function is a data consistency term that accounts for the model of \eqref{eq:degModel}.  The symbol $\nabla:\mathbb{R}^{\mathcal{M}_A\times \mathcal{N}_A}\rightarrow\mathbb{R}^{\mathcal{M}_A\times \mathcal{N}_A\times 2}$ represents the discrete gradient. 
Recall that $\mathbb{R}^{\mathcal{M}_A\times \mathcal{N}_A\times 2}$ is a vector space comprised of a three dimensional array.  Each pixel of a two-dimensional array yields a gradient vector with two elements in it: the horizontal component and the vertical component; the third dimension of the array to represent the different components: the first/second slice of this array is the horizontal/vertical component, respectively.
The $\|\cdot\|_{w,2}$ symbol represents a weighted $L_2$ norm, defined as
\begin{equation*}
  \|x\|_{w,2} = \left( w_1\,x_1^2 + w_2\,x_2^2 + \cdots + w_N\,x_N^2 \right)^{1/2}.
\end{equation*}

Each term in the objective function was divided by the relevant number of elements in the corresponding norm ($N_A$ or $N_M$) to make the regularization parameter robust to changing image sizes.  The data consistency term was multiplied by $1/2$ as a convenience so that the factor cancels when multiplied by $2$ during the computation of its derivative.  Note that one could simply multiply the entire objective function by $2$ and yield an equivalent problem.
Problem \eqref{eq:diChromProb} is a constrained least-squares problem.  The value of each pixel is constrained to be greater than $0$ (since the value represents a magnitude) and less than $1$ (a normalized maximum value).

Note that the gradient of the interpolated image and the proton image should not be related everywhere across the image.
As an extreme example, consider a region of the metabolic image without any hyperpolarized metabolite (which would appear dark in the metabolic image).
In this region, we want the values of the interpolated image to remain small (and not fluctuate with the gradient of the proton image).
More generally, we want the gradient to be similar for pixels where the values of the metabolic images are high, but unrelated for pixels where the values of the metabolic images are low.
To address this, we make use of the weighted norm in \eqref{eq:diChromProb}.  Similarly to the Morphology Enabled Dipole Inversion (MEDI) algorithm of \cite{liu2012morphology}, the weights are set to the values of the metabolic image linearly interpolated to be the size of the proton image.

If the user parameter $\lambda$ is very small, then the underdetermined linear system of \eqref{eq:degModel} dictates the output.  As $\lambda$ is increased, the regularization term $\mathcal{R}(I_M) = \frac{\lambda}{2\,N_H}\left\| \nabla I_M - \nabla I_A \right\|_{w,2}^2$ encourages the solution to have gradient vectors that are more and more similar to the gradient vectors of $A$, especially so where the values of $w$ are high.

\subsection{Solving the optimization problem}
The problem presented in \eqref{eq:diChromProb} is a convex optimization problem; thus, a solution can be determined with known algorithms and existing software solutions  \cite{grant2008_1,grant2008_2,diamond2016cvxpy}.  However, by forming an equivalent optimization problem, the interpolated image can be determined with the more efficient Fast Iterative Shrinkage Threshold Algorithm (FISTA).  This algorithm exhibits convergence of $\mathcal{O}(1/k^2)$ where $k$ is the iteration number, requiring fewer iterations than other methods for a given error.

FISTA solves problems of the form
\begin{equation}
  \underset{I_M}{\text{minimize}} \hspace{1em} F(I_M) + G(I_M).
  \label{eq:fistaProb}
\end{equation}
Let $F$ and $G$ be defined as follows:
\begin{align*}
  F(I_M) &= \frac{1}{2\,N_M} \| D\, B I_M - M \|_{Fr}^2 +
    \frac{\lambda}{2\,N_H}\left\| \nabla I_M - \nabla A \right\|_{w,2}^2 \text{ and } \\
  G(I_M) &= \mathbb{I}_{[0,1]}(I_M).
\end{align*}
Here, $\mathbb{I}_{[0,1]}$ is the indicator function of the $[0,1]$ interval applied to each element of the input individually.  It equals $0$ if every element is within the interval and infinity otherwise.  Then, problems \eqref{eq:diChromProb} and \eqref{eq:fistaProb} are equivalent (i.e. they are solved by the same solution set).

To improve the rate of convergence, we chose to use FISTA with line search \cite{scheinberg2014fast}.  A general description of this algorithm is presented in the appendix.  The proximal operator of $G$, required by the algorithm, is a Euclidean projection onto the $[0,1]^{N_H}$ set: $\text{prox}_{tG}(\cdot) = \min( \max( \cdot, 0 ), 1 )$ for any $t>0$, where the $\max$ and $\min$ operations are performed on each component of the input.

\subsection{Accounting for additional contrast}
\label{sec:intro:addtlContrast}

As we stated in section \ref{sec:intro}, the contrast of the high resolution image is necessarily weighted by proton density; this is unavoidable with standard MRI.  Therefore, there is only signal where there is tissue.  However, there may be additional contrasts imposed on the anatomical images such as $T_2$ or $T_1$ contrast.  If this contrast is the same as the contrast of the metabolite, then there is not an issue.  However, the contrast in the tissue of the anatomical image may be the negative of the contrast of the metabolic images, implying that the gradients of the anatomical image are opposite the desired gradients of the interpolated image.

It may be known, \textit{a priori}, how the contrast of the anatomical image relates to the metabolic image.  If that is the case, then the user should choose to either use $I_A$ or $-I_A$ as the anatomical reference if the contrast is the same as or opposite to the metabolic contrast, respectively.
If the relationship of the contrast (and the gradient directions) is not known \textit{a priori}, a simple metric can be used to determine whether to interpolate the metabolic images with $I_A$ or $-I_A$.  We chose to use both to interpolate the metabolic image, yielding $I_M^{(+)}$ and $I_M^{(-)}$, respectively.  Then, the result with contrast most similar to the contrast of the input metabolic image is selected as the final output.

\subsection{Di-chromatic interpolation}
Algorithm \ref{alg:diChromInterp} is the complete di-chromatic algorithm for interpolating MR data from a multi-slice acquisition.  The multiple slices may be acquisitions of adjacent slices in space, or it may be acquisitions of the same slice at different times.  It is assumed that the slices of the anatomical and metabolic volumes $\hat{V}_A$ and $\hat{V}_M$ are well registered.


\setlength{\textfloatsep}{1em}
\begin{algorithm}[ht]
  \protect
  \caption{Di-chromatic Interpolation \label{alg:diChromInterp}}

  \textbf{Inputs:} $\hat{V}_A$, $\hat{V}_M$, $B$, $D$, $\lambda$



  \textbf{For} each slice of $V_A$ and $V_M$ \textbf{\{}

    \hspace{1em} Set $\hat{A}$ and $\hat{M}$ to the current slice of $\hat{V}_A$ and $\hat{V}_M$, respectively.

    \hspace{1em} Set $A = \hat{A} / \max( \hat{A} )$ and $M = \hat{M} / \max( \hat{M} )$.

    \hspace{1em} Linearly interpolate $M$ to the size of $A$ to determine $w$.

    \hspace{1em} Solve problem \eqref{eq:diChromProb} with inputs $M$ and $A$ to determine $I_M^{(+)}$.

    \hspace{1em} Solve problem \eqref{eq:diChromProb} with inputs $M$ and $(1 - A)$ to determine $I_M^{(-)}$.

    \hspace{1em} Set the current slice of $\hat{V}_I^{(+)}$ and $\hat{V}_I^{(-)}$ to
      $I_M^{(+)}\cdot\max(\hat{M})$
      
    \hspace{2em} and $I_M^{(-)}\cdot\max(\hat{M})$, respectively.

    \hspace{1em} Store the value of $w\cdot\max(\hat{M})$ as a slice in a new volume $\hat{V}_w$.

  \textbf{\}}

  \textbf{If (} $\| \hat{V}_w - \hat{V}_I^{(+)} \|_{\hat{V}_w,2} < \| \hat{V}_w - \hat{V}_I^{(-)} \|_{\hat{V}_w,2}$ \textbf{)\{}

    \hspace{1em} $\hat{V}_I = \hat{V}_I^{(+)}$

  \textbf{\} Else \{}

      \hspace{1em} $\hat{V}_I = \hat{V}_I^{(-)}$

  \textbf{\}}

  \textbf{Output:} $\hat{V}_I$
\end{algorithm}

\subsection{Comparing to an existing algorithm}

An underlying idea for several of the existing algorithms is a regularization term that encourages edges of both images to be in the same location.  This is done with parallel level sets \cite{ehrhardt2016pet,schramm2017evaluation} or joint total variation (JTV) regularization \cite{chen2013calibrationless,mehranian2017synergistic}. (Recall that the $y-$level set of a function $f$ is the subset of its domain where the function equals $y$.)
For comparison purposes we have altered the method of \cite{mehranian2017synergistic}, which uses JTV, for reconstructed magnitude images as follows:
\begin{equation}
  \begin{aligned}
    \underset{ {I_A,I_M} }{ \text{minimize} } &\hspace{1em} \frac{1}{2\,N_A}\|I_A - A\|_{Fr}^2
      + \frac{1}{2\,N_M} \| D \, B \, I_M - M\|_{Fr}^2 + \\
      &\hspace{1em} + \frac{1}{N_A} \, \lambda \, \| ( \nabla I_A, \nabla I_M ) \|_{\text{JTV}}.
  \end{aligned}
  \label{eq:dichromInterp_stateOfTheArt}
\end{equation}
The $(\cdot,\cdot)$ notation represents array concatenation along the last dimension of the input arrays; therefore, $( \nabla I_A, \nabla I_M )\in\mathbb{R}^{\mathcal{M}_A\times \mathcal{N}_A\times 4}$.
The JTV norm is a group sparsity metric that computes the sum of the magnitudes of the $4$ element concatenated gradient vectors.

Problem \eqref{eq:dichromInterp_stateOfTheArt} can be solved with the Primal Dual Hybrid Gradient method \cite{chambolle2011first,esser2010general,pock2009algorithm}.  The resulting $I_A\in\mathbb{R}^{\mathcal{M}_A\times \mathcal{N}_A}$ is a denoised version of the anatomical image \cite{rudin1992nonlinear} and the resulting $I_M\in\mathbb{R}^{\mathcal{M}_M\times \mathcal{N}_M}$ is the interpolated metabolic image.
We will compare the method presented in this paper, detailed in section \ref{sec:methods}, to the JTV interpolation of solving \eqref{eq:dichromInterp_stateOfTheArt}.

\section{Experiments}

In addition to a numerical phantom, MR images of an experimental phantom, humans, and a pig were processed for this paper.  Images were collected with a $3$ Tesla General Electric MR750 clinical scanner (GE Healthcare, Waukesha, WI).
For all studies, hyperpolarized [1-$^{13}$C] pyruvate was generated in a $5$ Tesla SPINlab polarizer operating at $0.8$ Kelvin.  Samples were polarized for at least $2.5$ hours and then rapidly dissolved and neutralized.  All data processed in this manuscript were originally acquired for other purposes.

\subsection{Numerical phantom}

To demonstrate the utility of this technique, we created a set of phantom tumors (homogeneous and heterogeneous) in a metabolic image, shown in Fig. \ref{fig:phantomTumors}a.  We used the Shepp-Logan phantom to represent the anatomy (Fig. \ref{fig:phantomTumors}b).  The tumors were super-imposed on the Shepp-Logan phantom in Fig. \ref{fig:phantomTumors}c.  Tumor 1 is homogeneous and one side corresponds to an edge in the anatomical image, tumor 2 is heterogeneous and is mismatched with the edge in the underlying anatomy, tumor 3 is heterogeneous and in a homogeneous region of anatomy, and tumor 4 is homogeneous with edges that match the underlying anatomy.  We performed linear and di-chromatic interpolation and compare the results in Fig. \ref{fig:phantomTumors}g.


\subsection{Experimental phantom}

Three solution filled bottles were used for an experimental phantom.  A 5 mL 13C-urea(1 mol) and a 5 mL 13C-acetate(1 mol) phantom were placed along with a 5 mL saline phantom.  All cylinder phantoms were simultaneously imaged in an axial plane with three separate acquisitions.  For the proton imaging, a stock CINE SSFP sequence was used.  
For the carbon imaging, a spectral-spatial (SPSP) acquisition and single-shot spiral read-out were used.
Imaging specifics were: $80$ Hz single band excitation, flip angles for lactate/pyruvate were $90^\circ / 8^\circ$, the field of view was $30 \times 30$ $\text{cm}^2$, recon matrix = $128 \times 128$, real pixels size = $10$ mm, and slice thickness = $20$ mm.

\subsection{Human acquisitions}

Prior to injection of the hyperpolarized solution, the pH, pyruvate and residual paramagnetic agent concentration, polarization, and temperature were measured.  After release by the pharmacist, a $0.43$ mL/kg dose of approximately $250$ mM pyruvate was injected at a rate of $5$ mL/s, followed by a $20$ mL saline flush.

\subsection{Porcine cardiac acquisitions}

Metabolic images were acquired of a Danish domestic feed pig weighing 40 kg with a clamp shell transmit coil and a $16$ channel array receive coil (Rapid Biomedical, Rimpar, Germany).  A $25$ mL of approximately $180$ mM pyruvate solution was injected $20$ seconds after dissolution into central venous access over $10$ seconds with a $15$ mL saline flush.
The pig received intravenous propofol (12 mg initial dose; thereafter 0.4 mg/kg/h for maintenance anaesthesia), intravenous fentanyl ($8$ $\mu$g/kg/h), and was mechanically ventilated.  Catheterization was performed through the femoral vein for the administration of hyperpolarized [$1$‐$^{13}$C] pyruvate.  Imaging was done in the supine position.

Proton CINE cardiac short-axis images were acquired with a stock GE Healthcare provided FIESTA sequence with cardiac gating and breath-hold. Imaging specifics were: flip angle = $55^\circ$, field of view = $400 \times 400$, recon matrix = $512 \times 512$, real pixels size = $2.2$ mm, and slice thickness = $10$ mm.
Pyruvate cardiac short-axis images were obtained using a spectral-spatial excitation and a single-shot spiral read out with cardiac gating. Imaging specifics were: $80$ Hz single band excitation, flip angles for lactate/pyruvate were $90^\circ / 8^\circ$, the field of view was $30 \times 30$ $\text{cm}^2$, recon matrix = $128 \times 128$, real pixels size = $10$ mm, slice thickness = $20$ mm.

\subsubsection{Human cardiac images}
\label{sec:cardiacExp}
Proton density weighted images of a healthy volunteer were acquired using a multi-slice free-breathing gradient echo sequence with a $3\times 3$ mm$^2$ in-plane resolution, an echo time of $2.8$ ms, and a repetition time of $12.8$ ms with the system's body coil.
This data was collected using a commercial software package (RTHawk, HeartVista, Los Altos, CA).
The pyruvate, lactate, and bicarbonate images were acquired alternately using a multi-slice free-breathing cardiac-gated sequence with a $12.5\times 12.5$ mm$^2$ in-plane resolution and a field-of-view of $75\times 75$ cm$^2$.  For the metabolic images, a Helmholz clamshell transmit coil and an $8$-channel paddle receive array were used.
A single band spectral-spatial excitation scheme was used with a single-shot spiral readout trajectory \cite{cunningham2008pulse}; the flip angles for pyruvate, lactate, and bicarbonate were $20^\circ$, $30^\circ$, and $30^\circ$, respectively \cite{lau2011spectral,tang2019regional}.
Bolus tracking was used to trigger the acquisition with real-time frequency and power calibration \cite{tang2019regional}.

\subsubsection{Human brain}

One hyperpolarized brain dataset was acquired in a healthy volunteer with a variable-resolution single-shot echo-planar Imaging acquisition \cite{gordon2020variable} using a birdcage coil for transmit with an integrated $24$ element receiver (Rapid Biomedical, Würzburg, Germany). Scan parameters were $125$ ms TR, $30.7$ ms TE, $32 \times 32$ matrix size, eight $1.5$ cm slices with an axial orientation. Pyruvate was excited with a $20^\circ$ flip angle and acquired at $7.5\times 7.5$ mm$^2$ resolution, while lactate and bicarbonate received a $30^\circ$ flip angle and were acquired at $15\times 15$ mm$^2$ in-plane resolution \cite{lau2011spectral}. Data acquisition started $5$ seconds after the end of saline injection. Twenty frames were acquired with a $3$ second temporal resolution, yielding a total scan time of one minute.  For anatomic reference, a 3D Inversion Recovery Spoiled Gradient Echo sequence dataset was acquired with a birdcage transmit coil and an integrated $8$ channel receive coil \cite{gordon2017development}.  Scan parameters were $6.7$ ms TR, $2.5$ ms TE, $450$ ms IR time, $25.6\times 25.6\times 18.6$ cm$^2$ field of view, $256\times 256\times 124$ matrix ($1\times 1\times 1.5$ mm$^3$ resolution).

\subsubsection{Human prostate}
\label{sec:prostateExp}
Data of a prostate with cancer were acquired with a three-dimensional undersampled spectroscopic imaging sequence with compressed sensing reconstruction \cite{larson2011fast,chen2018technique}.
Whole organ coverage was achieved with a resolution of $8\times8\times8$ mm$^3$ and a spectral bandwidth of $540$ Hz.
Metabolite volumes were acquired every $2$ seconds using varying flip angles to improve the signal-to-noise ratio \cite{xing2013optimal}.
For anatomic reference, $T_2$ weighted proton images were acquired with a repetition time of $6$ seconds, an echo time of $102$ ms, a field of view of $18\times 18$ cm$^2$, an image size of $384\times 384$, and a slice thickness of $3$ mm.

\section{Results}
\label{sec:results}

In this section, we present results from a healthy porcine heart, a human heart of a healthy volunteer, a human brain from a healthy volunteer, and a human prostate with cancer.  Finally, we present results showing the effect of changing the regularization parameter.




\subsection{Numerical phantom}

\begin{figure}[H]
  \begin{center}
    \includegraphics[width=0.95\linewidth]{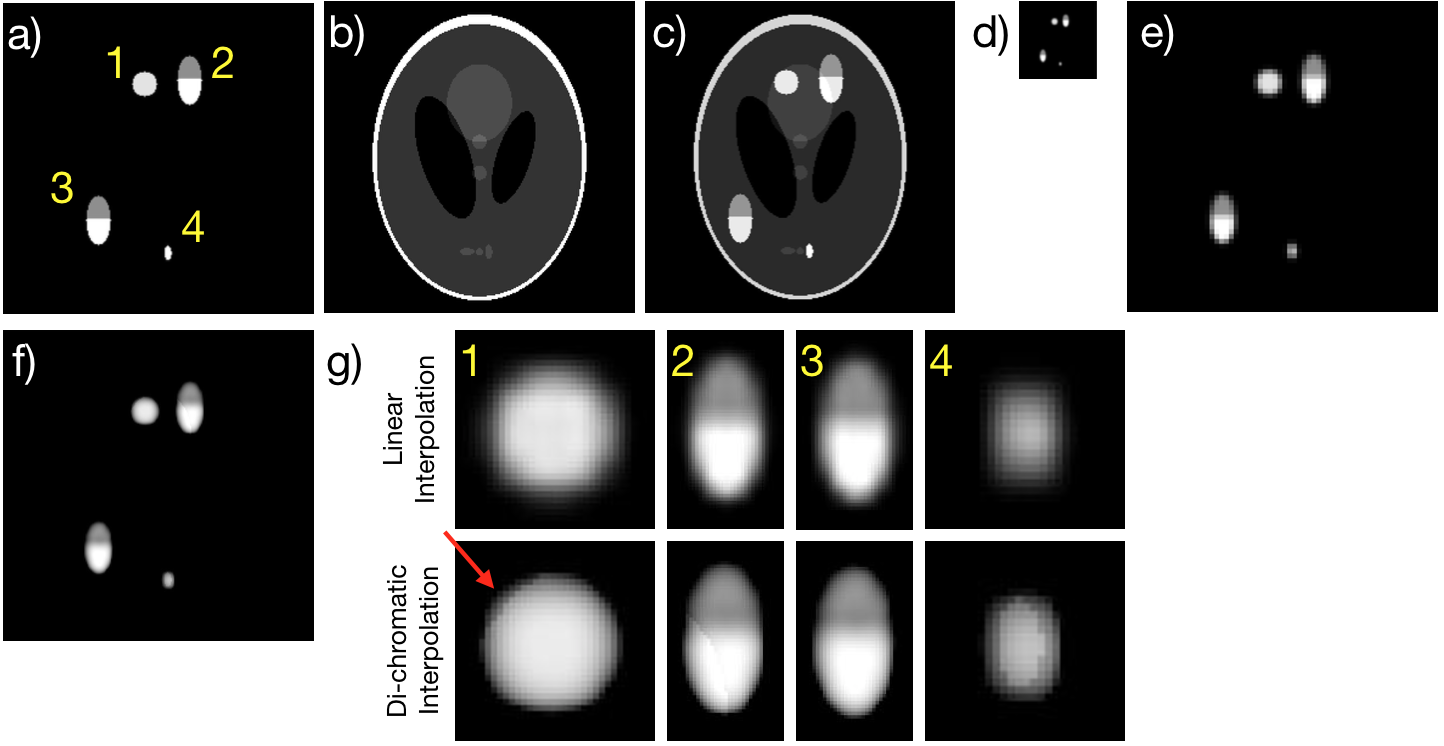}
  \end{center}
  \caption{ \label{fig:phantomTumors} a) A high resolution metabolic numerical phantom made of two homogeneous tumors (labeled `1' and `4') and two heterogeneous tumors (labeled `2' and `3').  b) The Shepp-Logan phantom will serve as the high-resolution numerical anatomical phantom.  c) The tumors super-imposed on the anatomical phantom.  d)  The metabolic phantom reduced in resolution to $25\%$ of its original size.  e) The reduced resolution phantom of d interpolated to the size of the phantom with nearest neighbor interpolation.  f) The result of di-chromatic interpolation on the low resolution metabolic phantom.  g) Sub-images of the tumors resulting from linear interpolation (top) and di-chromatic interpolation (bottom). The red arrow indicates the portion of the tumor that was aligned with an anatomical edge.}
\end{figure}

The result of di-chromatic interpolation applied to the numerical phantom is shown in Fig. \ref{fig:phantomTumors}.  The relative errors of the linear interpolation and the di-chromatic interpolation results are $0.33$ and $0.31$, which quantifies an improvement with the di-chromatic interpolation.  Note that relative error is defined as $ \|\text{estimate} - \text{true}\|_{\text{Fr}} / \| \text{true} \|_{\text{Fr}}$ where $\|\cdot\|_{\text{Fr}}$ denotes the Frobenius norm.

In the di-chromatic interpolated image, tumor 1 becomes sharper on the side that corresponds to the anatomical edge (as indicated by the red arrow) and the other sides do not significantly change quality.  In tumor 2, the mismatched anatomical edge is incorporated into the interpolated image, but the overall contrast and presence of the tumor is not significantly affected.  The incorporated edge empowers the observer to localize the tumor in the anatomy well.  Tumor 3 does not significantly change in the di-chromatic interpolated result; its contrast is maintained.  Tumor 4 becomes sharper by incorporating anatomical edges.  Both heterogeneous phantoms retain their structure and contrast, even though they are not present in the underlying anatomic image used for interpolation.  In all cases, di-chromatic interpolation offers a benefit.

\subsection{Experimental Phantom}

\begin{figure}[H]
  \begin{center}
    \includegraphics[width=0.7\linewidth]{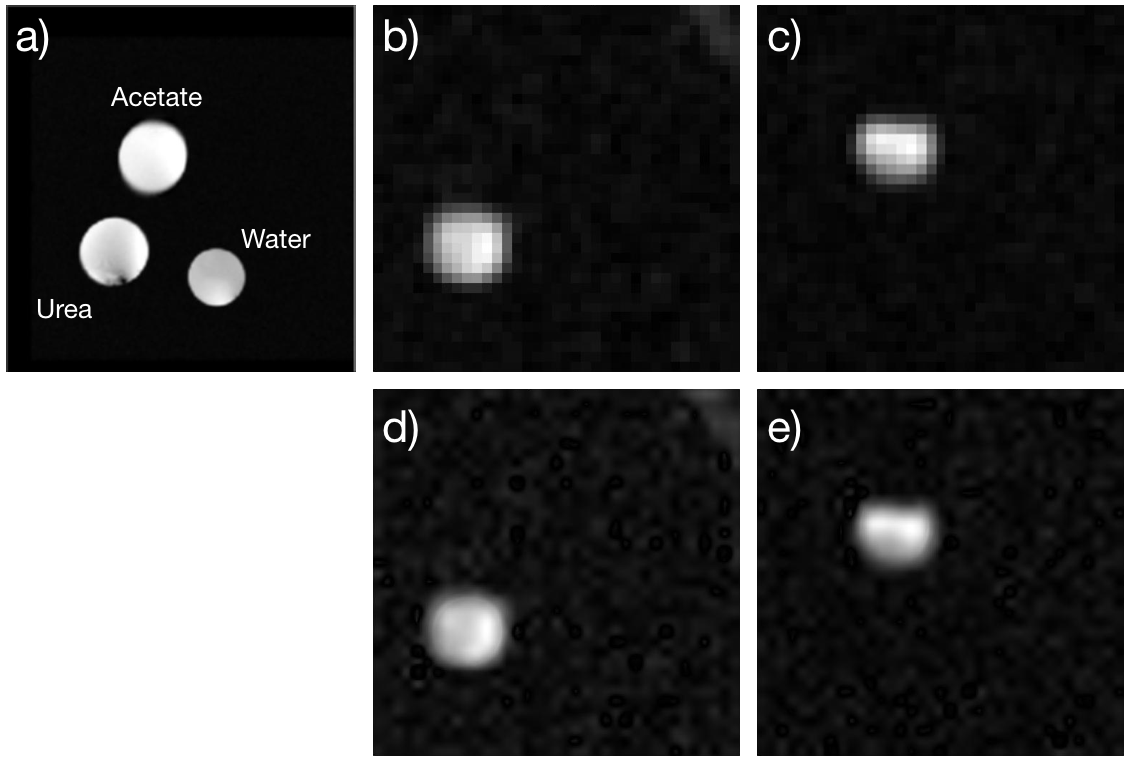}
  \end{center}
  \caption{ \label{fig:expPhantom} 
  Data and results from the experimental phantom: (a) proton image, (b) native resolution urea image, (c) native resolution acetate image, (d) interpolated urea image, and (e) interpolated acetate image.}
\end{figure}

Figure \ref{fig:expPhantom} shows the original images of the experimental phantom along with the di-chromatic interpolated metabolic images.  Note the clear separation of the different metabolites; and note that the saline phantom without a carbon substrate does not have significant signal in either of the interpolated metabolic images.  The flat top of the acetate phantom is likely due to a susceptibility-induced artifact resulting from the spiral acquisition.  As a result, the di-chromatic interpolation generates strong edges on the left, right, and bottom of the carbon phantoms and reflects the lack of a distinct edge near the tops of the bottles.  This is similar to the results shown with tumor 1 in the numerical phantom of Fig. \ref{fig:phantomTumors}.

\subsection{Porcine cardiac images}

\begin{figure}[H]
  \begin{center}
    \includegraphics[width=0.8\linewidth]{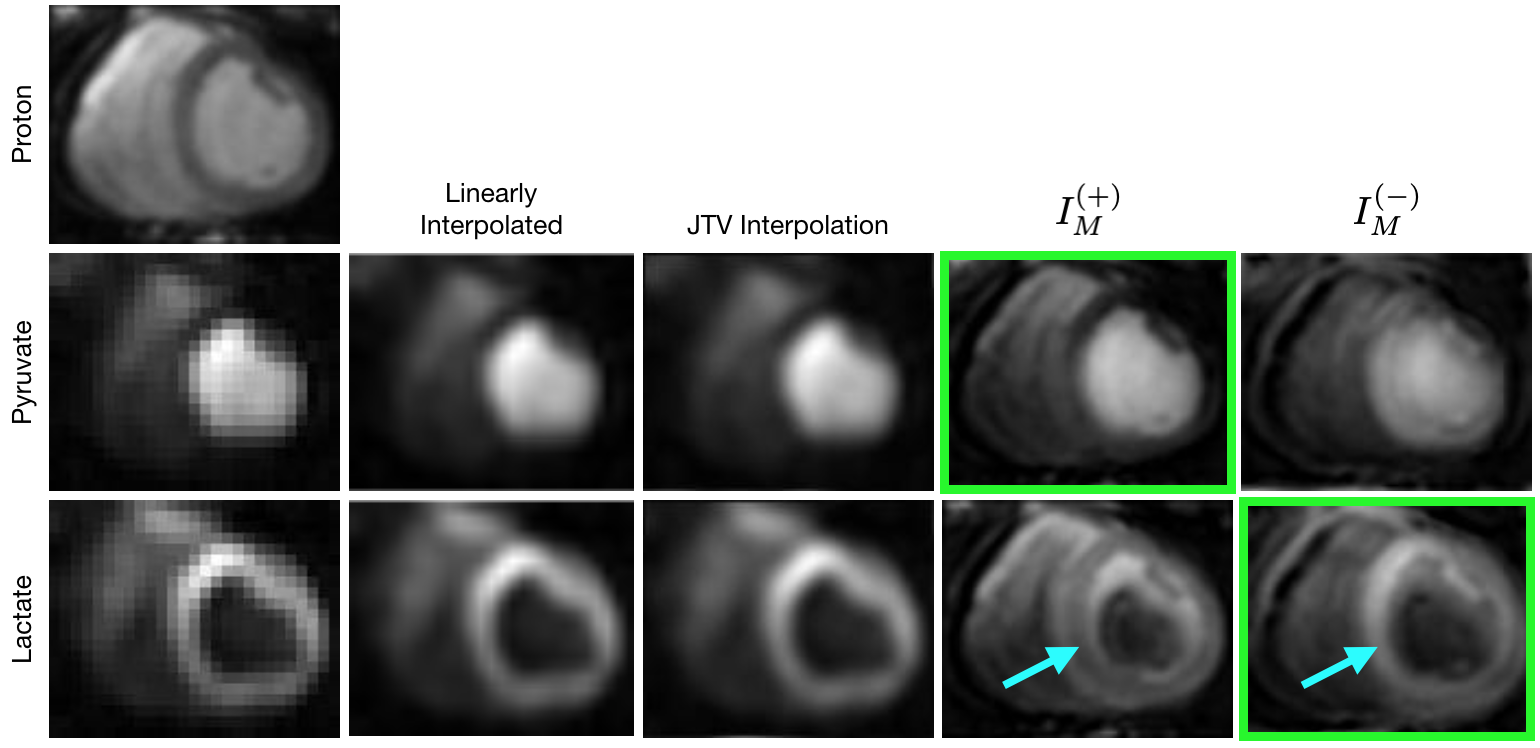}
  \end{center}
  \caption{ \label{fig:pigHeartDiChromContrasts} (Row-1) a high resolution proton image of a porcine heart.  (Row-2/Row3) The metabolic images of pyruvate and lactate, respectively, at the same slice location.  (Column-1) Metabolic images are increased in size to that of the proton image with nearest-neighbor interpolation.  (Column-2) The result of linearly interpolating the metabolic images to the size of the proton image.  (Column-3) The result of using JTV interpolation on the metabolic images; note that the some of the artifacts remaining in the linearly interpolated images have been eliminated and the JTV interpolation result appears smoother.  (Column-4/Column-5) $I_M^{(+)}$ and $I_M^{(-)}$ from Alg. \ref{alg:diChromInterp}, respectively, with a regularization parameter $\lambda=10$.  The cyan arrows indicate a location where the contrast of $I_M^{(+)}$ is opposite that of $I_M^{(-)}$.  The green boxes show the contrast selected by the di-chromatic interpolation algorithm for final output; note that $I_M^{(+)}$ was selected as the output for pyruvate but $I_M^{(-)}$ was selected as the output for lactate.}
\end{figure}

Figure \ref{fig:pigHeartDiChromContrasts} shows results of imaging protons, pyruvate, and lactate in the porcine heart.  In this case, the contrast of the lactate is opposite that of the proton image.  Note the high intensity of the lactate in the myocardium and the low local contrast of the same region in the proton image.  The green boxes show the interpolated image selected by Algorithm \ref{alg:diChromInterp}, which presents the contrast most similar to the non-interpolated image.

The interpolated images show the pyruvate primarily localized to the blood pools and lactate primarily localized to the myocardium, which matches with our expectation since pyruvate is the injected substrate and lactate would be produced in the muscle.  In this example, the papillary muscle is also delineated in the proton image, and the interpolation shows relatively little pyruvate and high lactate in this structure, as expected.

Figure \ref{fig:pigHeartDiChromContrasts} also compares the di-chromatic interpolation algorithm to linear interpolation and JTV interpolation.  JTV interpolation eliminates some amount of the artifacts that remain with linear interpolation, but the results are very similar.  Since JTV interpolation is much more computationally intensive, it is not necessarily the case that the improvement in quality is worth the cost.  The di-chromatic interpolation results incorporate the anatomical information much more.

\subsection{Human cardiac images}

\begin{figure}[H]
  \begin{center}
    \includegraphics[width=0.8\linewidth]{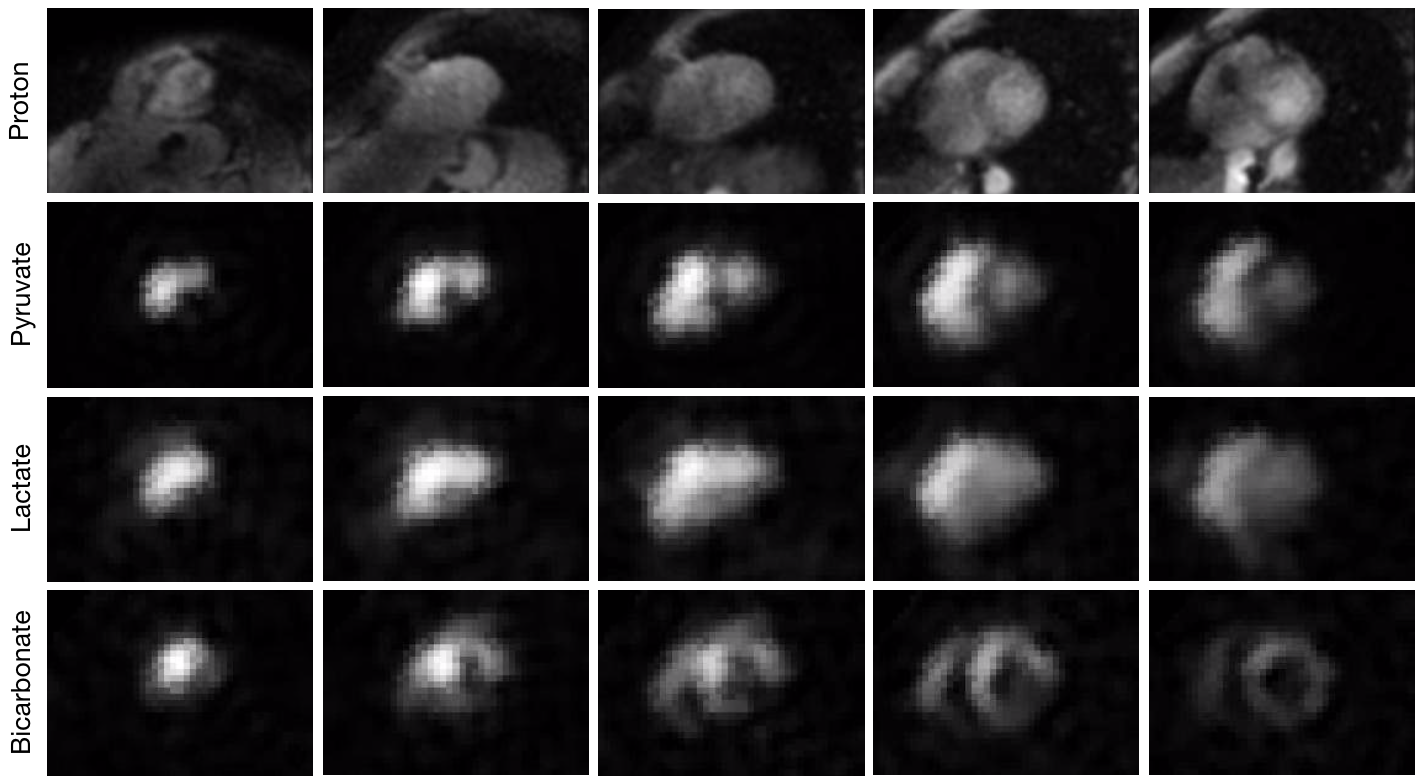}
  \end{center}
  \caption{ \label{fig:humanHeartOrig} High-resolution proton image and low-resolution metabolic image of the cardiac study.  The first/second/third/fourth row shows proton/pyruvate/lactate/bicarbonate, respectively.  There are five different slices of the heart shown in the five different columns. }
\end{figure}

\begin{figure}[H]
  \begin{center}
    \includegraphics[width=0.8\linewidth]{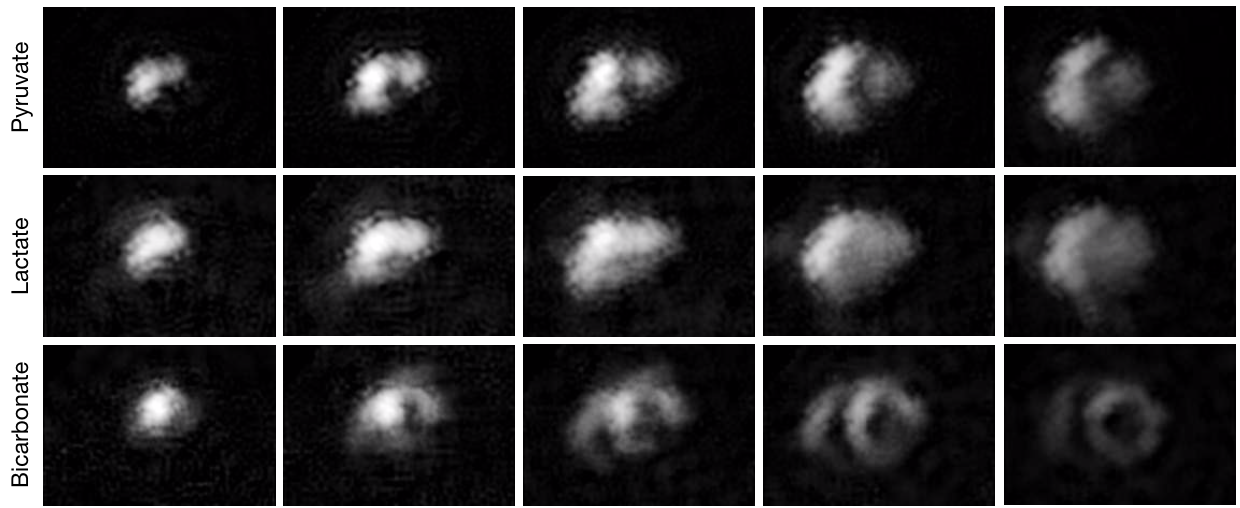}
  \end{center}
  \caption{ \label{fig:humanHeartDiChromInterp} Interpolated metabolic image of the cardiac study with $\lambda=1$.  The first/second/third row shows pyruvate/lactate/bicarbonate, respectively.  There are five different slices of the heart shown in the five different columns. }
\end{figure}

Figures \ref{fig:humanHeartOrig} shows five slices of a heart at a single point in time where much of the pyruvate has been converted to lactate and bicarbonate through metabolism at its native resolution; Fig. \ref{fig:humanHeartDiChromInterp} shows the interpolated image.
Note the additional detail in the interpolated image.  The bicarbonate image retains a markedly different spatial distribution compared to the pyruvate and lactate images.  In particular, the bicarbonate is accurately localized to the myocardium.

\begin{figure}[H]
  \begin{center}
    \includegraphics[width=0.8\linewidth]{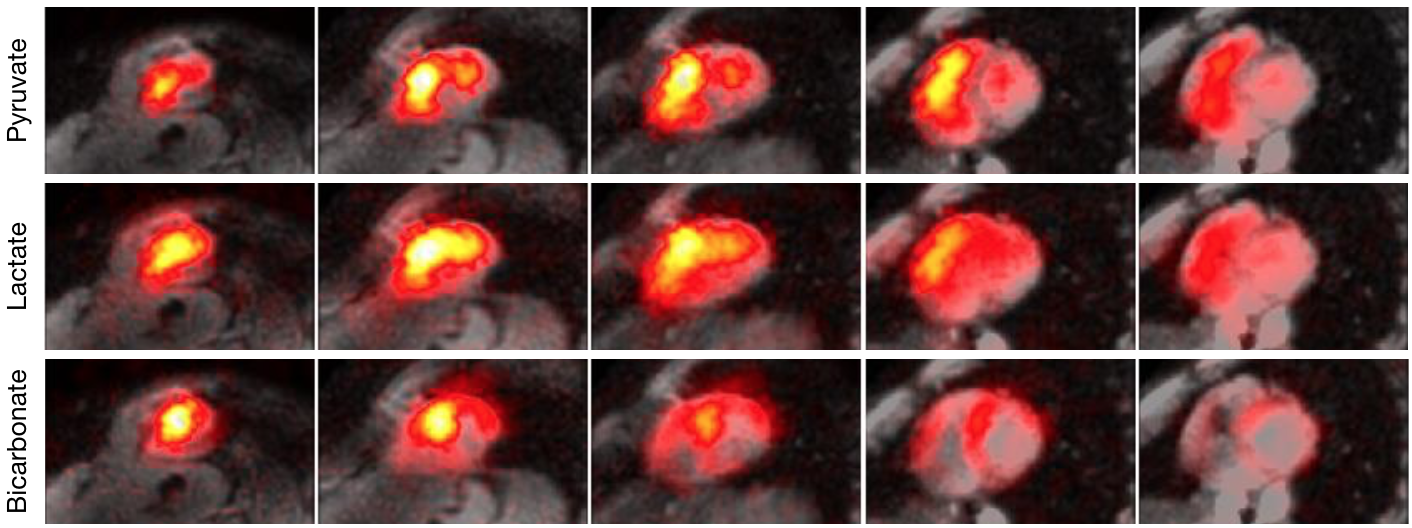}
  \end{center}
  \caption{ \label{fig:humanHeartDiChromFused} Interpolated metabolic image of the cardiac study with $\lambda=1$ fused with the proton image using the CLS Fusion algorithm.  The first/second/third row shows pyruvate/lactate/bicarbonate, respectively.  There are five different slices of the heart shown in the five different columns. }
\end{figure}

An alternative method of presenting the metabolic image is to fuse it with the proton image.  Figure \ref{fig:humanHeartDiChromFused} shows the interpolated image of Fig. \ref{fig:humanHeartDiChromInterp} made into a false color image with Matlab's hot colormap and fused with the proton image using the CLS fusion algorithm \cite{dwork2017formulation}.  The CLS fusion algorithm  was designed to present the anatomical information while retaining the information in the color image.

\subsection{Human brain}

\begin{figure}[H]
  \begin{center}
    \includegraphics[width=0.8\linewidth]{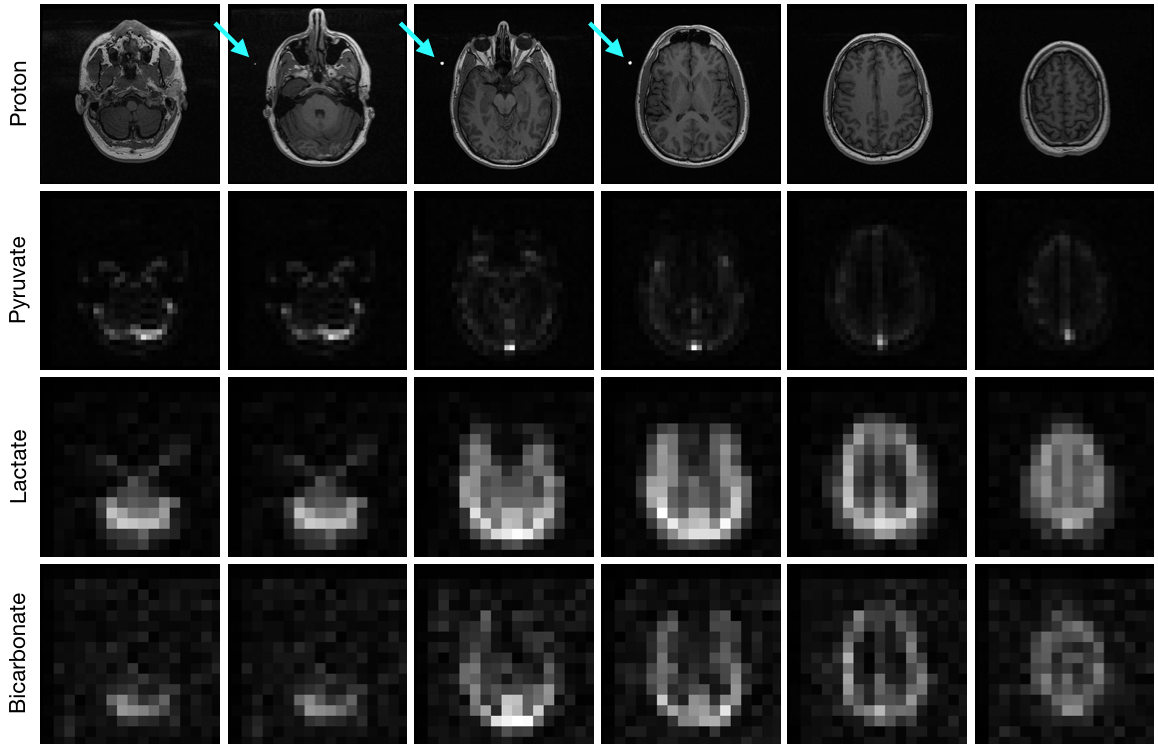}
  \end{center}
  \caption{ \label{fig:brainOrig} (Row-1) a high resolution proton image.  (Row-2/Row-3/Row-4) The metabolic images of pyruvate, lactate, and bicarbonate, respectively.  (Columns 1-6) Individual slices of the brain.  Note that the native resolution of the pyruvate is twice that of the lactate and bicarbonate.  The cyan arrows point to a urea phantom used for power calibration. }
\end{figure}

Figure \ref{fig:brainOrig} shows the high resolution proton images and the lower resolution pyruvate, lactate, and bicarbonate images for $6$ slices of the brain.  In order to improve the signal-to-noise ratio of the metabolites with lower abundance, lactate and bicarbonate, the resolution of those images was reduced by a factor of $2$ from that of pyruvate \cite{gordon2020variable}.

\begin{figure}[H]
  \begin{center}
    \includegraphics[width=0.8\linewidth]{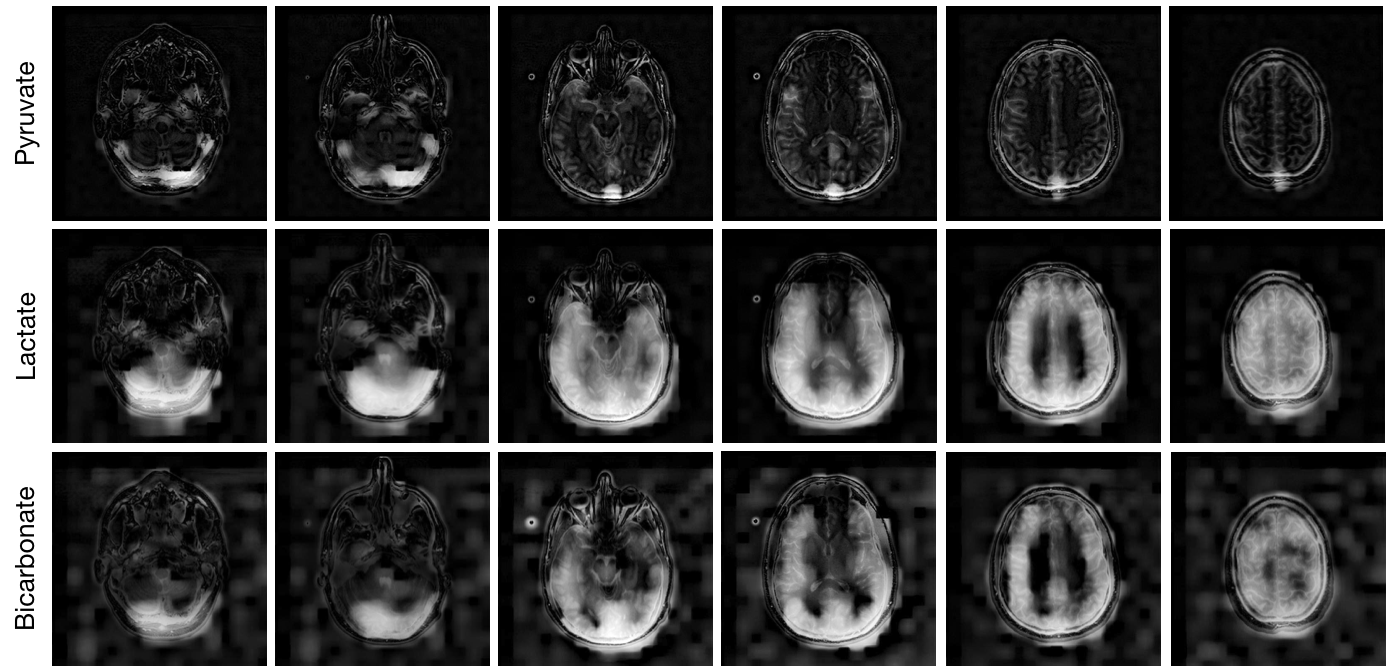}
  \end{center}
  \caption{ \label{fig:brainInterped} Di-chromatic interpolated image with $\lambda=10$ of (Row-1/Row-2/Row-3) pyruvate, lactate, and bicarbonate, respectively.  (Columns 1-6) Individual slices of the brain.  Note that the native resolution of the pyruvate is twice that of the lactate and bicarbonate.}
\end{figure} 

Figure \ref{fig:brainInterped} shows the di-chromatic interpolated images of the brain for pyruvate, lactate, and bicarbonate.  
The interpolation puts the majority of metabolite signals within the brain as expected.  However, some of the fine structure contrast observed (e.g. metabolite signals outside the brain or in the CSF) do not reflect metabolic activity.  With the enhanced details in the interpolated metabolite images, it is easier for the observer to determine the regions of the brain where the activity is taking place based on the metabolic images alone.

\subsection{Human prostate}

\begin{figure}[H]
  \begin{center}
    \includegraphics[width=0.8\linewidth]{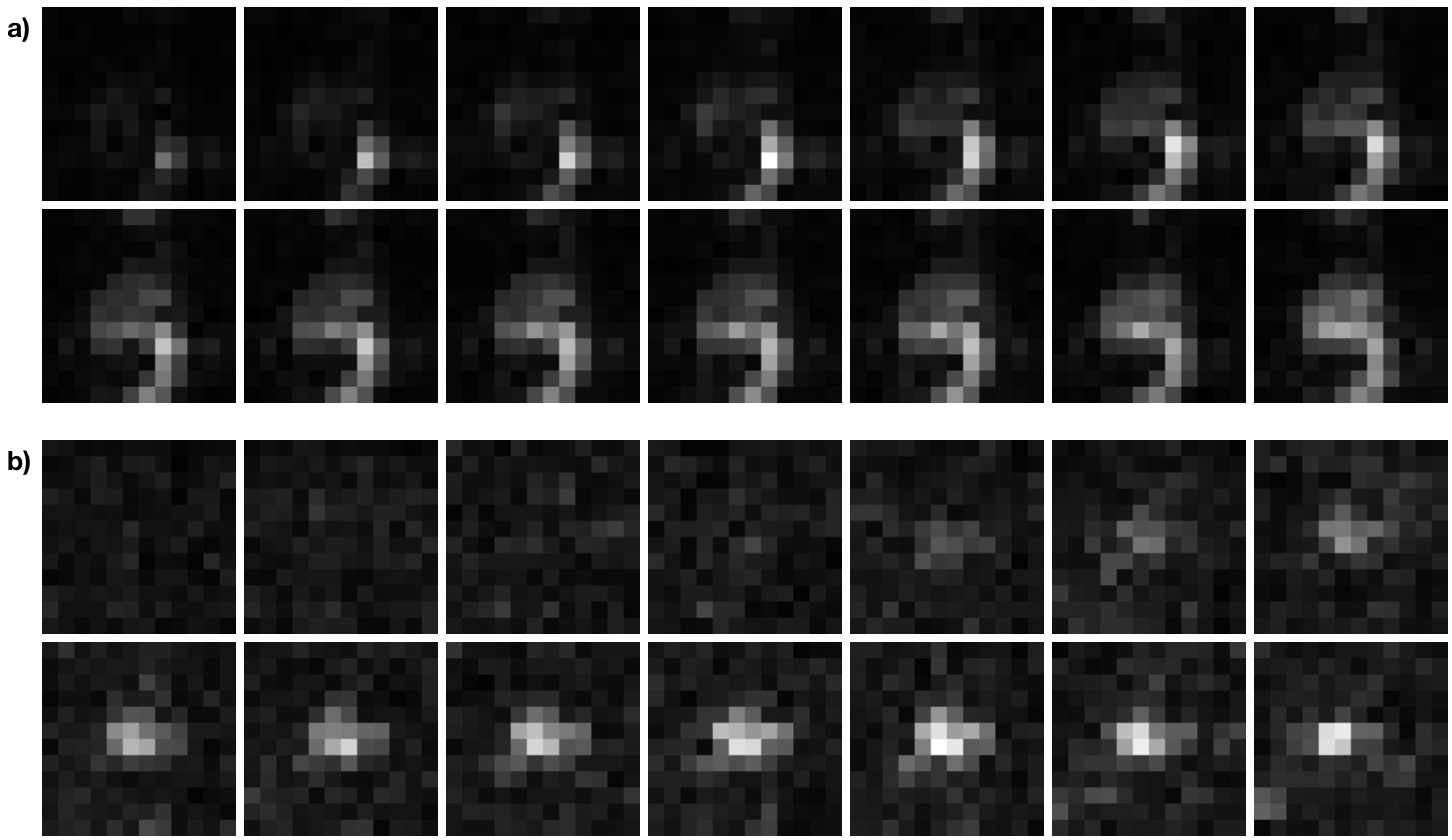}
  \end{center}
  \caption{ \label{fig:origProstate} Dynamic images from a single slice of a) pyruvate and b) lactate in the prostate at native ($8\times 8$ mm$^2$) resolution, ordered temporally  from left-to-right and then top-to-bottom.  Each image shows the pyruvate content at a different time; images are separated by approximately 2 seconds. }
\end{figure}

\begin{figure}[H]
  \begin{center}
    \includegraphics[width=0.8\linewidth]{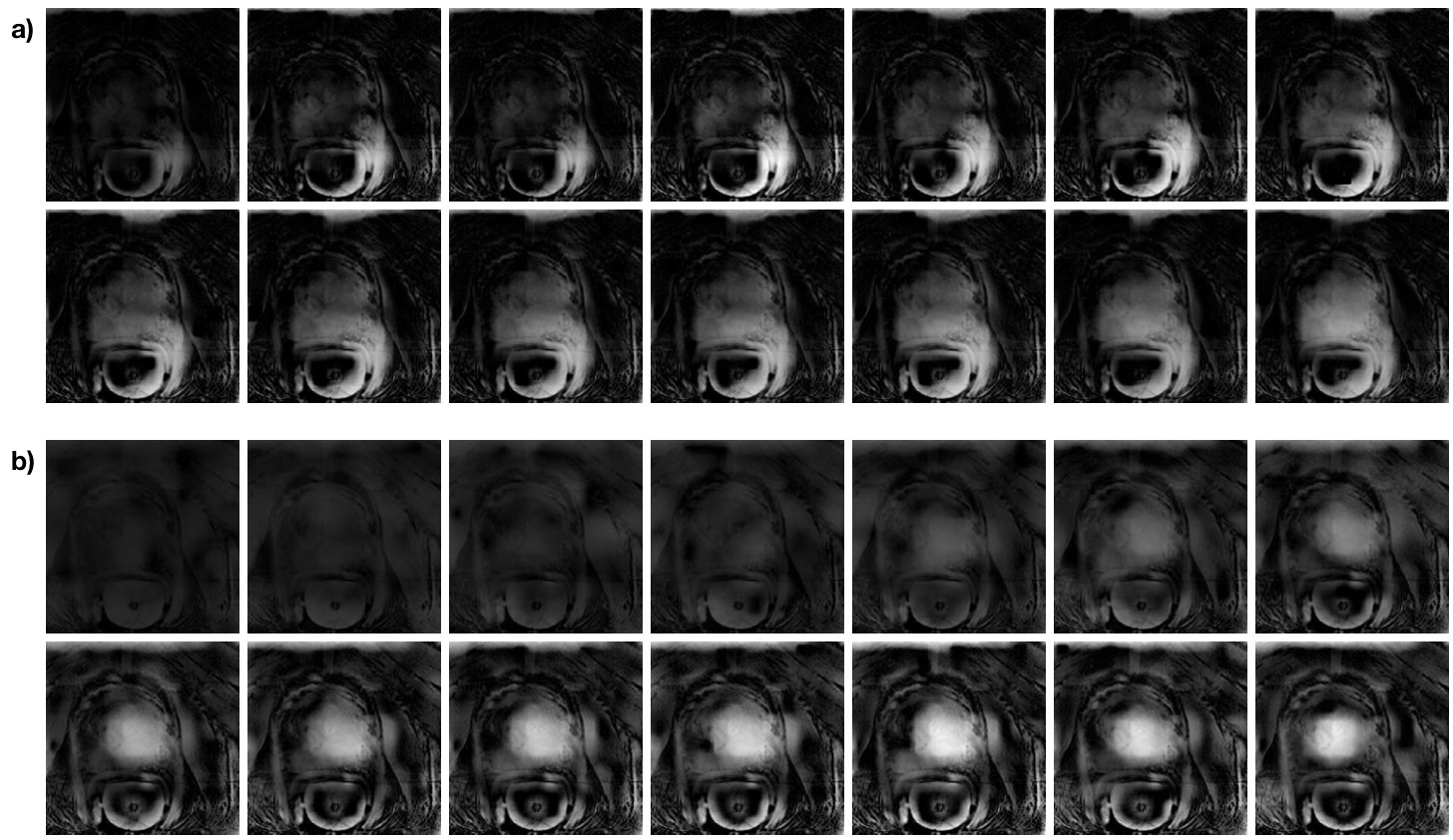}
  \end{center}
  \caption{ \label{fig:prostateDiChromInterp} Interpolated images of a) pyruvate and b) lactate in the prostate with $\lambda=10$, ordered temporally  from left-to-right and then top-to-bottom.  Each image shows the pyruvate content at a different time; images are separated by approximately 2 seconds. }
\end{figure}

Figure \ref{fig:origProstate} shows pyruvate and lactate images of a slice of a prostate taken approximately $2$ seconds apart in their native resolution.  The Di-chromatic Interpolation Algorithm \ref{alg:diChromInterp} with a regularization parameter of $\lambda=10$ was applied with the anatomical reference image presented in \ref{fig:sampleData} (left).  The results are shown in \ref{fig:prostateDiChromInterp}.  Note that one is able to better comprehend where in the prostate the metabolic activity is taking place by only looking at the interpolated metabolic image.

Although the image is more detailed, Fig. \ref{fig:prostateDiChromInterp} shows a limitation of the di-chromatic interpolation algorithm.  The pyruvate largely arrives at this slice of the prostate through a blood vessel and diffuses into the surrounding tissue.  We would not expect the regions of fat to have a large pyruvate uptake.  The di-chromatic interpolation algorithm does not take this physiology into account.  So, though the interpolation better localizes the metabolic energy to regions of tissue, the fat region encompassed with the red circle in Fig. \ref{fig:sampleData} is bright.

\subsection{The regularization Parameter}

\begin{figure}[H]
  \begin{center}
    \includegraphics[width=0.7\linewidth]{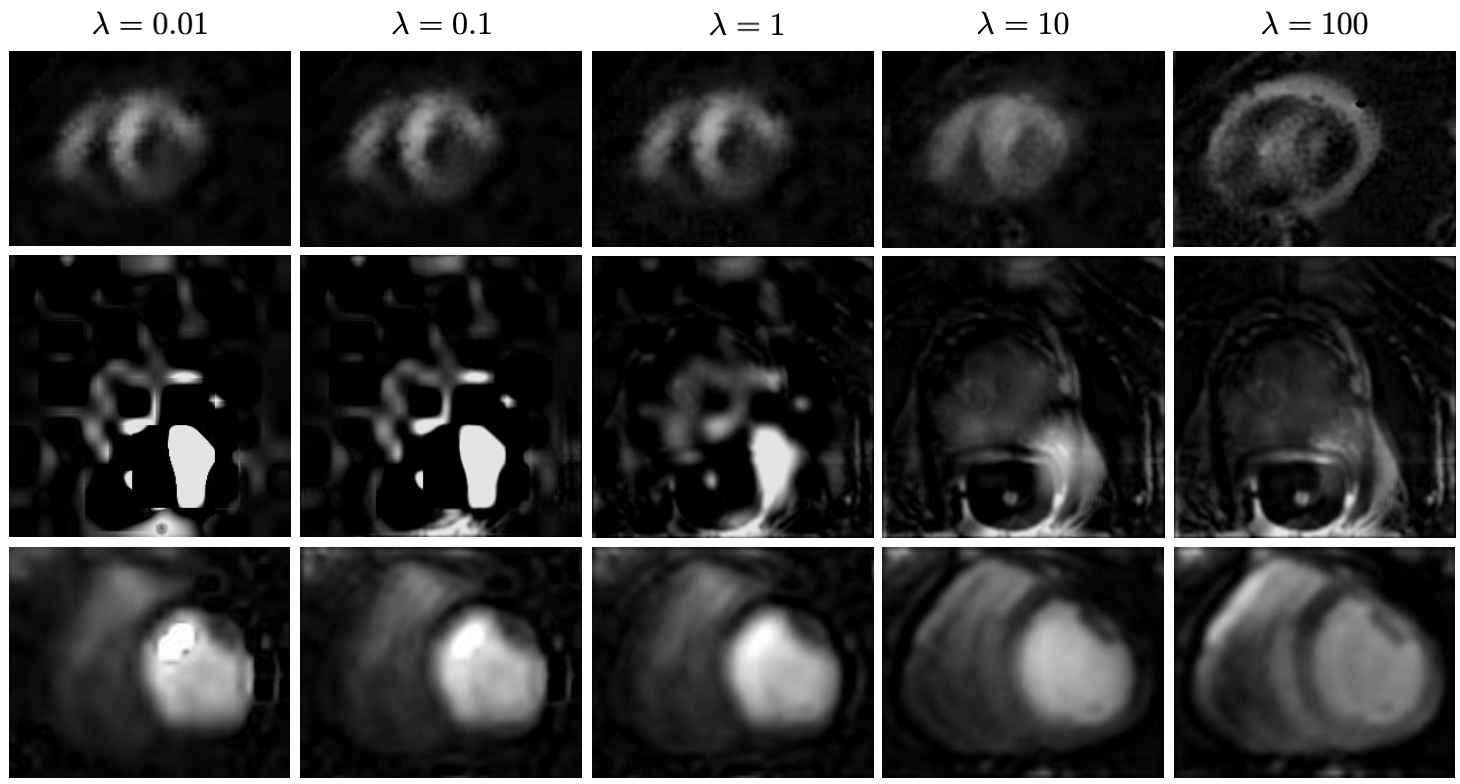}
  \end{center}
  \caption{ \label{fig:changingLambdas} Di-chromatic interpolation results for various values of the regularization parameter $\lambda$.  The value of $\lambda$ (from left to right) is $10^{-2}$, $10^{-1}$, $1$, $10^1$, and $10^2$.  Each row shows how the results change for data of bicarbonate in a human heart (top row, Fig. \ref{fig:humanHeartOrig}), pyruvate in a human prostate (middle row, Fig \ref{fig:origProstate}), and pyruvate in a porcine heart (bottom row, Fig. \ref{fig:pigHeartDiChromContrasts}).
  With low values of $\lambda$, the regularization term has little effect and the data consistency attempts to de-blur the image, which gathers energy in the images.  This effect can be seen dramatically in the prostate with $\lambda=10^{-2}$ and $\lambda=10^{-1}$ }
\end{figure}

The regularization variable $\lambda$ is a user-selected parameter.
Figure \ref{fig:changingLambdas} shows how the details of the interpolated image are altered by changes to the regularization parameter of \eqref{eq:diChromProb}.  The first / second / third rows of Fig. \ref{fig:changingLambdas} show results for data from the human heart / prostate / and porcine heart, respectively.  Generally, as the regularization parameter increases, the interpolated image appears more similar to the proton image.  This can be seen in the pyruvate image of the porcine heart with $\lambda=100$; the image appears extremely similar to the corresponding proton image (seen in Fig. \ref{fig:prostateDiChromInterp}) and the metabolic information has almost entirely been lost.  As the regularization parameter is reduced, the interpolated image appears less natural, becoming artificially detailed to better solve the data-consistency term in the objective function of \eqref{eq:diChromProb}.  The data consistency attempts to de-blur the image, which gathers energy in the images.  This is especially notable in the prostate.  Note the contrasts of the underlying proton image changes as $\lambda$ is increased for the bicarbonate image of the heart.  Whereas $I_M^{(+)}$ is selected as the final output for most values of $\lambda$, in order to maintain a contrast closest to the original un-interpolated image, $I_M^{(-)}$ is selected for the final output when $\lambda=100$.

\section{Discussion}
The di-chromatic interpolation algorithm improves the observer's comprehension of the location of metabolic activity attained from an injection of a hyperpolarized solution.
This has great potential for improved localization of tumors and malfunctions of cardiac tissue.
Additionally, the method performs a deblurring operation which gathers the energy of the hyperpolarized metabolite according to the blur kernel.
This reduces the apparent spread of the activity, possibly reducing the possibility of falsely mischaracterizing tissue as metabolically hyperactive.

Note that the interpolation algorithm does not generate the same results that would be achieved with a high resolution image (were it possible to create one).  The interpolated image is a better reflection of the metabolic activity in the anatomy, but not necessarily a completely physical one.
Our results suggest that this method is well suited for cardiac imaging because the high resolution anatomical structures match the expected compartmentalization of the hyperpolarized metabolites.  The proton images clearly define the blood and myocardium, and these are surrounded by lung tissue, while the hyperpolarized metabolites typically are distributed in the blood and/or myocardium, but not in the lungs.  

However, in the brain and prostate we observed that the high resolution anatomical contrast was likely too complex and detailed compared to the expected hyperpolarized metabolite distributions.  While we observed some apparent improvements in localization of metabolites to the prostate and brain, the algorithm also places metabolite signals in other structures such as intra-abdominal and subcutaneous fat where it is very unlikely that the metabolites resides.  The results for some applications may be improved by selecting the contrast of the high resolution anatomical image acquired with the metabolic images to better match the expected distribution of hyperpolarized metabolites.

The regularization parameter determines the quality of the output to a significant degree.  If it is too low, then the output is largely determined by the noise in the input.  If it is too high, then the output becomes too similar to the anatomical image.  In our experience, though, we have found that a single regularization parameter works well for all slices of the same volume.  Our hope is that future collections can depend on quality results from the same regularization parameter determined for data of the same resolution.  We have not demonstrated this yet.  Future investigations could identify a heuristic that automatically selects the regularization parameter, perhaps by finding the parameter at the knee of the pareto-optimal curve \cite{das1999characterizing}.

With dynamic data (images of metabolic processing through time) the temporal results are correlated.  This has been taken into account in MR image reconstruction with total-variation regularization imposed in the temporal dimension \cite{wang2017real}.  In a similar vein, total-variation regularization can be added to problem \eqref{eq:diChromProb} for dynamic data.

Gordon et al. have shown that the low signal-to-noise ratio of some hyperpolarized compounds (e.g. bicarbonate) can be compensated with a lower spatial resolution \cite{gordon2020variable}.  For data collected in this way, it may be beneficial to use the pyruvate image as the reference high-resolution image to interpolate the lower-resolution bicarbonate image.

Though we presented this technique in the context of hyperpolarized MRI, this technique may be applicable to other imaging systems.
For example, it may be used to combine data from an MR / Positron Emission Tomography (PET) machine.  Or, in another MR application, it may be used to interpolate xenon pulmonary images \cite{mugler1997mr}.  When inhaled, the xenon gas perfuses into regions where tissue is absent.  Since the di-chromatic interpolation may select the negative of the proton MR image, the result could be informative.
We leave the investigation of these possibilities for future work.

\section{Conclusion}
\label{sec:conclusion}
In this work, we present the di-chromatic interpolation algorithm for MRI that informs the interpolation of a low-resolution image of hyperpolarized compounds using the gradients of a high-resolution image.
The algorithm is based on known physics, and the solution space is further limited with a heuristic method incorporating information from a high-resolution anatomical image.
We show results using data of the human prostate, the human heart, and a porcine heart.  We demonstrate the interpolation algorithm for data that varies spatially and data that varies temporally.  This algorithm accepts, as input, reconstructed images.  Since it does not require raw data collected by the scanner, it is feasible to deploy this algorithm for use on clinical patient data (for data that will be collected in the future or that has been collected in the past).

\section*{Compliance with Ethical Standards}
ND has received post-doctoral training funding from the American Heart Association (grant number 20POST35200152).
ND has received funding from the Quantitative Biosciences Institute at UCSF (no grant number).
JG has received funding from the National Institute of Health / National Institute of Biomedical Imaging and Bioengineering (grant number U01EB026412).
PL has received funding from the the National Institute of Health (grant number NIH R01 HL136965).

No conflicts of interest, financial or otherwise, are declared by the authors.

All procedures performed in studies involving human participants were in accordance with the ethical standards of the institutional and/or national research committee and with the 1964 Helsinki declaration and its later amendments or comparable ethical standards.
MR data of humans was gathered with Institutional Review Board (IRB) approval and Health Insurance Portability and Accountability Act (HIPAA) compliance.  Informed consent was obtained from all individual participants included in the study.

All applicable international, national, and/or institutional guidelines for the care and use of animals were followed.
Animal experiments were done in accordance with relevant laws and ethics under permission from The Animal Experiments Inspectorate of Denmark.

\section*{Acknowledgments}
The authors would like to thank Roselle Abraham, Rahul Aggarwal, Robert Bok, Hsin-Yu Chen, John Kurhanewicz, James Slater, and Daniel Vigneron for their assistance in the imaging of human subjects.
The authors would like to thank Gennifer T. Smith for her helpful suggestions regarding the editing of this document.
ND would like to thank the Quantitative Biosciences Institute at UCSF and the American Heart Association as funding sources for this work.

\appendix
\label{sec:appendixFista}
\section{Fast Iterative Shrinkage Threshold Algorithm}

The Fast Iterative Shrinkage Threshold Algorithm (FISTA) solves problems of the form
\begin{equation*}
    \underset{x\in\mathbb{R}^N}{\text{minimize}} \hspace{1em} F(x) + G(x),
\end{equation*}
where $F$ is differentiable and $G$ has a simple proximal operator \cite{beck2009fast,scheinberg2014fast}.
The FISTA algorithm with line search is described in algorithm \ref{alg:fistaLS}.
Note that $\langle\cdot,\cdot\rangle$ represents an inner product.
To initialize the algorithm, set $v^{(0)} = x^{(0)}$, where $x^{(0)}$ is the initial guess and can be any value.
Select a $t_0>0$, and select a maximum number of iterations $K$.
Select a backtracking line search parameter $r\in(0,1)$ (a common choice of $r$ is $0.9$) and select a step size scaling parameter $s>1$ (a common choice of $s$ is $1.25$).

\begin{algorithm}[H]
  \caption{FISTA with line search \protect\label{alg:fistaLS}}

    \textbf{For} $k=1,2,\ldots,K$

    \hspace{1em} $t_k = s\,t_{k-1}$
    
    \hspace{1em} \textbf{While} true

    \hspace{2em} \textbf{If} $k==1$
    
    \hspace{3em} $\theta_k = 1$
    
    \hspace{2em} \textbf{Else}
    
    \hspace{3em} $\theta_k = \text{positive root of } t_{k-1}\,\theta^2 =t_k\,\theta_{k-1}^2(1-\theta)$

    \hspace{2em} \textbf{End If}

    \hspace{2em} $y^{(k)} = (1-\theta_k)x^{(k-1)} + \theta_k v^{(k-1)}$

    \hspace{2em} $x^{(k)} = \prox_{t_k\,G}\left(y^{(k)}-t_k\,\nabla F(y^{(k)})\right)$
    
    \hspace{2em} \textbf{If} $F(x^{(k)}) \leq F(y^{(k)}) + \left\langle \nabla F(y^{(k)}), x^{(k)}-y^{(k)} \right\rangle + 
      \frac{1}{2t}\|x^{(k)}-y^{(k)}\|_2^2$

    \hspace{3em} \textbf{break}

    \hspace{2em} \textbf{End If}

    \hspace{2em} $t_k := r \, t_k$

    \hspace{1em} \textbf{End While}

    \hspace{1em} $v^{(k)} = x^{(k-1)} + \frac{1}{\theta_k}\left(x^{(k)}-x^{(k-1)}\right)$
    
    \textbf{End For}

\end{algorithm}


\begin{thebibliography}{10}

\bibitem{ardenkjaer2003increase}
Jan~H Ardenkj{\ae}r-Larsen, Bj{\"o}rn Fridlund, Andreas Gram, Georg Hansson,
  Lennart Hansson, Mathilde~H Lerche, Rolf Servin, Mikkel Thaning, and Klaes
  Golman.
\newblock Increase in signal-to-noise ratio of $> 10,000$ times in liquid-state
  {NMR}.
\newblock {\em Proceedings of the National Academy of Sciences},
  100(18):10158--10163, 2003.

\bibitem{golman2006real}
Klaes Golman, Mikkel Thaning, et~al.
\newblock Real-time metabolic imaging.
\newblock {\em Proceedings of the National Academy of Sciences},
  103(30):11270--11275, 2006.

\bibitem{schroeder2011hyperpolarized}
Marie~A Schroeder, Kieran Clarke, Stefan Neubauer, and Damian~J Tyler.
\newblock Hyperpolarized magnetic resonance: a novel technique for the in vivo
  assessment of cardiovascular disease.
\newblock {\em Circulation}, 124(14):1580--1594, 2011.

\bibitem{nelson2013metabolic}
Sarah~J Nelson, John Kurhanewicz, Daniel~B Vigneron, Peder~EZ Larson, Andrea~L
  Harzstark, Marcus Ferrone, Mark van Criekinge, Jose~W Chang, Robert Bok,
  Ilwoo Park, et~al.
\newblock Metabolic imaging of patients with prostate cancer using
  hyperpolarized [1-13{C}] pyruvate.
\newblock {\em Science translational medicine}, 5(198):198ra108--198ra108,
  2013.

\bibitem{kurhanewicz2019hyperpolarized}
John Kurhanewicz, Daniel~B Vigneron, Jan~Henrik Ardenkjaer-Larsen, James~A
  Bankson, Kevin Brindle, Charles~H Cunningham, Ferdia~A Gallagher, Kayvan~R
  Keshari, Andreas Kjaer, Christoffer Laustsen, et~al.
\newblock Hyperpolarized {13C} {MRI}: path to clinical translation in oncology.
\newblock {\em Neoplasia}, 21(1):1--16, 2019.

\bibitem{grist2019quantifying}
James~T Grist, Mary~A McLean, Frank Riemer, Rolf~F Schulte, Surrin~S Deen,
  Fulvio Zaccagna, Ramona Woitek, Charlie~J Daniels, Joshua~D Kaggie, Tomasz
  Matys, et~al.
\newblock Quantifying normal human brain metabolism using hyperpolarized
  [{1-13C}] pyruvate and magnetic resonance imaging.
\newblock {\em NeuroImage}, 189:171--179, 2019.

\bibitem{cunningham2016hyperpolarized}
Charles~H Cunningham, Justin~YC Lau, Albert~P Chen, Benjamin~J Geraghty,
  William~J Perks, Idan Roifman, Graham~A Wright, and Kim~A Connelly.
\newblock Hyperpolarized {13C} metabolic {MRI} of the human heart: initial
  experience.
\newblock {\em Circulation research}, 119(11):1177--1182, 2016.

\bibitem{larson2018investigation}
Peder~EZ Larson, Hsin-Yu Chen, Jeremy~W Gordon, Natalie Korn, John Maidens,
  Murat Arcak, Shuyu Tang, Mark Criekinge, Lucas Carvajal, Daniele Mammoli,
  et~al.
\newblock Investigation of analysis methods for hyperpolarized 13{C}-pyruvate
  metabolic {MRI} in prostate cancer patients.
\newblock {\em {NMR} in Biomedicine}, 31(11):e3997, 2018.

\bibitem{golman2008cardiac}
Klaes Golman, J~Stefan Petersson, Peter Magnusson, Edvin Johansson, Per
  {\AA}keson, Chun-Ming Chai, Georg Hansson, and Sven M{\aa}nsson.
\newblock Cardiac metabolism measured noninvasively by hyperpolarized 13{C}
  {MRI}.
\newblock {\em Magnetic Resonance in Medicine: An Official Journal of the
  International Society for Magnetic Resonance in Medicine}, 59(5):1005--1013,
  2008.

\bibitem{wang2019hyperpolarized}
Zhen~J Wang, Michael~A Ohliger, Peder~EZ Larson, Jeremy~W Gordon, Robert~A Bok,
  James Slater, Javier~E Villanueva-Meyer, Christopher~P Hess, John
  Kurhanewicz, and Daniel~B Vigneron.
\newblock Hyperpolarized 13c mri: State of the art and future directions.
\newblock {\em Radiology}, 291(2):273--284, 2019.

\bibitem{knoll2016joint}
Florian Knoll, Martin Holler, Thomas Koesters, Ricardo Otazo, Kristian Bredies,
  and Daniel~K Sodickson.
\newblock Joint {MR}-{PET} reconstruction using a multi-channel image
  regularizer.
\newblock {\em IEEE transactions on medical imaging}, 36(1):1--16, 2016.

\bibitem{cui2018wavelet}
Xuelin Cui, Lamine Mili, Ge~Wang, and Hengyong Yu.
\newblock Wavelet-based joint {CT}-{MRI} reconstruction.
\newblock {\em Journal of {X}-ray science and technology}, 26(3):379--393,
  2018.

\bibitem{ehman2017pet}
Eric~C Ehman, Geoffrey~B Johnson, Javier~E Villanueva-Meyer, Soonmee Cha,
  Andrew~Palmera Leynes, Peder Eric~Zufall Larson, and Thomas~A Hope.
\newblock Pet/mri: where might it replace pet/ct?
\newblock {\em Journal of Magnetic Resonance Imaging}, 46(5):1247--1262, 2017.

\bibitem{fessler2010model}
Jeffrey~A Fessler.
\newblock Model-based image reconstruction for {MRI}.
\newblock {\em {IEEE} signal processing magazine}, 27(4):81--89, 2010.

\bibitem{boyd2004convex}
Stephen Boyd, Stephen~P Boyd, and Lieven Vandenberghe.
\newblock {\em Convex optimization}.
\newblock Cambridge university press, 2004.

\bibitem{li2009fusion}
Zhenhua Li and Henry Leung.
\newblock Fusion of multispectral and panchromatic images using a
  restoration-based method.
\newblock {\em IEEE transactions on geoscience and remote sensing},
  47(5):1482--1491, 2009.

\bibitem{garzelli2016review}
Andrea Garzelli.
\newblock A review of image fusion algorithms based on the super-resolution
  paradigm.
\newblock {\em Remote Sensing}, 8(10):797, 2016.

\bibitem{wu2019remote}
Honglin Wu, Shuzhen Zhao, Jianming Zhang, and Chaoquan Lu.
\newblock Remote sensing image sharpening by integrating multispectral image
  super-resolution and convolutional sparse representation fusion.
\newblock {\em IEEE Access}, 7:46562--46574, 2019.

\bibitem{lukin2004high}
Alexey Lukin and Denis Kubasov.
\newblock High-quality algorithm for bayer pattern interpolation.
\newblock {\em Programming and Computer Software}, 30(6):347--358, 2004.

\bibitem{liu2012morphology}
Jing Liu, Tian Liu, Ludovic de~Rochefort, James Ledoux, Ildar Khalidov, Weiwei
  Chen, A~John Tsiouris, Cynthia Wisnieff, Pascal Spincemaille, Martin~R
  Prince, et~al.
\newblock Morphology enabled dipole inversion for quantitative susceptibility
  mapping using structural consistency between the magnitude image and the
  susceptibility map.
\newblock {\em Neuroimage}, 59(3):2560--2568, 2012.

\bibitem{grant2008_1}
Michael Grant and Stephen Boyd.
\newblock {CVX}: {Matlab} software for disciplined convex programming, version
  2.1.
\newblock \url{http://cvxr.com/cvx}, March 2014.

\bibitem{grant2008_2}
Michael Grant and Stephen Boyd.
\newblock Graph implementations for nonsmooth convex programs.
\newblock In V.~Blondel, S.~Boyd, and H.~Kimura, editors, {\em Recent Advances
  in Learning and Control}, Lecture Notes in Control and Information Sciences,
  pages 95--110. Springer-Verlag Limited, 2008.
\newblock \url{http://stanford.edu/~boyd/graph_dcp.html}.

\bibitem{diamond2016cvxpy}
Steven Diamond and Stephen Boyd.
\newblock {CVXPY}: A python-embedded modeling language for convex optimization.
\newblock {\em The Journal of Machine Learning Research}, 17(1):2909--2913,
  2016.

\bibitem{scheinberg2014fast}
Katya Scheinberg, Donald Goldfarb, and Xi~Bai.
\newblock Fast first-order methods for composite convex optimization with
  backtracking.
\newblock {\em Foundations of Computational Mathematics}, 14(3):389--417, 2014.

\bibitem{ehrhardt2016pet}
Matthias~J Ehrhardt, Pawel Markiewicz, Maria Liljeroth, Anna Barnes, Ville
  Kolehmainen, John~S Duncan, Luis Pizarro, David Atkinson, Brian~F Hutton,
  Sebastien Ourselin, et~al.
\newblock {PET} reconstruction with an anatomical {MRI} prior using parallel
  level sets.
\newblock {\em IEEE transactions on medical imaging}, 35(9):2189--2199, 2016.

\bibitem{schramm2017evaluation}
Georg Schramm, Martin Holler, Ahmadreza Rezaei, Kathleen Vunckx, Florian Knoll,
  Kristian Bredies, Fernando Boada, and Johan Nuyts.
\newblock Evaluation of parallel level sets and bowsher’s method as
  segmentation-free anatomical priors for time-of-flight {PET} reconstruction.
\newblock {\em IEEE transactions on medical imaging}, 37(2):590--603, 2017.

\bibitem{chen2013calibrationless}
Chen Chen, Yeqing Li, and Junzhou Huang.
\newblock Calibrationless parallel {MRI} with joint total variation
  regularization.
\newblock In {\em International Conference on Medical Image Computing and
  Computer-Assisted Intervention}, pages 106--114. Springer, 2013.

\bibitem{mehranian2017synergistic}
Abolfazl Mehranian, Martin~A Belzunce, Claudia Prieto, Alexander Hammers, and
  Andrew~J Reader.
\newblock Synergistic {PET} and {SENSE} {MR} image reconstruction using joint
  sparsity regularization.
\newblock {\em {IEEE} transactions on medical imaging}, 37(1):20--34, 2017.

\bibitem{chambolle2011first}
Antonin Chambolle and Thomas Pock.
\newblock A first-order primal-dual algorithm for convex problems with
  applications to imaging.
\newblock {\em Journal of mathematical imaging and vision}, 40(1):120--145,
  2011.

\bibitem{esser2010general}
Ernie Esser, Xiaoqun Zhang, and Tony~F Chan.
\newblock A general framework for a class of first order primal-dual algorithms
  for convex optimization in imaging science.
\newblock {\em {SIAM} Journal on Imaging Sciences}, 3(4):1015--1046, 2010.

\bibitem{pock2009algorithm}
Thomas Pock, Daniel Cremers, Horst Bischof, and Antonin Chambolle.
\newblock An algorithm for minimizing the mumford-shah functional.
\newblock In {\em 2009 {IEEE} 12th International Conference on Computer
  Vision}, pages 1133--1140. IEEE, 2009.

\bibitem{rudin1992nonlinear}
Leonid~I Rudin, Stanley Osher, and Emad Fatemi.
\newblock Nonlinear total variation based noise removal algorithms.
\newblock {\em Physica D: nonlinear phenomena}, 60(1-4):259--268, 1992.

\bibitem{cunningham2008pulse}
Charles~H Cunningham, Albert~P Chen, Michael Lustig, Brian~A Hargreaves, Janine
  Lupo, Duan Xu, John Kurhanewicz, Ralph~E Hurd, John~M Pauly, Sarah~J Nelson,
  et~al.
\newblock Pulse sequence for dynamic volumetric imaging of hyperpolarized
  metabolic products.
\newblock {\em Journal of magnetic resonance}, 193(1):139--146, 2008.

\bibitem{lau2011spectral}
Angus~Z Lau, Albert~P Chen, Ralph~E Hurd, and Charles~H Cunningham.
\newblock Spectral--spatial excitation for rapid imaging of {DNP} compounds.
\newblock {\em {NMR} in Biomedicine}, 24(8):988--996, 2011.

\bibitem{tang2019regional}
Shuyu Tang, Eugene Milshteyn, Galen Reed, Jeremy Gordon, Robert Bok, Xucheng
  Zhu, Zihan Zhu, Daniel~B Vigneron, and Peder~EZ Larson.
\newblock A regional bolus tracking and real-time {B1} calibration method for
  hyperpolarized {13C} {MRI}.
\newblock {\em Magnetic resonance in medicine}, 81(2):839--851, 2019.

\bibitem{gordon2020variable}
Jeremy~W Gordon, Adam~W Autry, Shuyu Tang, Jasmine~Y Graham, Robert~A Bok,
  Xucheng Zhu, Javier~E Villanueva-Meyer, Yan Li, Michael~A Ohilger,
  Maria~Roselle Abraham, et~al.
\newblock A variable resolution approach for improved acquisition of
  hyperpolarized {13C} metabolic {MRI}.
\newblock {\em Magnetic Resonance in Medicine}, 2020.

\bibitem{gordon2017development}
Jeremy~W Gordon, Daniel~B Vigneron, and Peder~EZ Larson.
\newblock Development of a symmetric echo planar imaging framework for clinical
  translation of rapid dynamic hyperpolarized {13C} imaging.
\newblock {\em Magnetic resonance in medicine}, 77(2):826--832, 2017.

\bibitem{larson2011fast}
Peder~EZ Larson, Simon Hu, Michael Lustig, Adam~B Kerr, Sarah~J Nelson, John
  Kurhanewicz, John~M Pauly, and Daniel~B Vigneron.
\newblock Fast dynamic {3D} {MR} spectroscopic imaging with compressed sensing
  and multiband excitation pulses for hyperpolarized 13c studies.
\newblock {\em Magnetic resonance in medicine}, 65(3):610--619, 2011.

\bibitem{chen2018technique}
Hsin-Yu Chen, Peder~EZ Larson, Jeremy~W Gordon, Robert~A Bok, Marcus Ferrone,
  Mark van Criekinge, Lucas Carvajal, Peng Cao, John~M Pauly, Adam~B Kerr,
  et~al.
\newblock Technique development of {3D} dynamic {CS-EPSI} for hyperpolarized
  {13C} pyruvate {MR} molecular imaging of human prostate cancer.
\newblock {\em Magnetic resonance in medicine}, 80(5):2062--2072, 2018.

\bibitem{xing2013optimal}
Yan Xing, Galen~D Reed, John~M Pauly, Adam~B Kerr, and Peder~EZ Larson.
\newblock Optimal variable flip angle schemes for dynamic acquisition of
  exchanging hyperpolarized substrates.
\newblock {\em Journal of magnetic resonance}, 234:75--81, 2013.

\bibitem{dwork2017formulation}
Nicholas Dwork, Eric~M Lasry, John~M Pauly, and Jorge Balb{\'a}s.
\newblock Formulation of image fusion as a constrained least squares
  optimization problem.
\newblock {\em Journal of Medical Imaging}, 4(1):014003, 2017.

\bibitem{das1999characterizing}
Indraneel Das.
\newblock On characterizing the ``knee" of the pareto curve based on
  normal-boundary intersection.
\newblock {\em Structural optimization}, 18(2-3):107--115, 1999.

\bibitem{wang2017real}
Yang Wang, Ning Cao, Zuojun Liu, and Yudong Zhang.
\newblock Real-time dynamic {MRI} using parallel dictionary learning and
  dynamic total variation.
\newblock {\em Neurocomputing}, 238:410--419, 2017.

\bibitem{mugler1997mr}
John~P Mugler~III, Bastiaan Driehuys, James~R Brookeman, Gordon~D Cates,
  Stuart~S Berr, Robert~G Bryant, Thomas~M Daniel, Eduard~E De~Lange, J~Hunter
  Downs, Christopher~J Erickson, et~al.
\newblock {MR} imaging and spectroscopy using hyperpolarized 129 {Xe} gas:
  preliminary human results.
\newblock {\em Magnetic resonance in medicine}, 37(6):809--815, 1997.

\bibitem{beck2009fast}
Amir Beck and Marc Teboulle.
\newblock A fast iterative shrinkage-thresholding algorithm for linear inverse
  problems.
\newblock {\em {SIAM} journal on imaging sciences}, 2(1):183--202, 2009.

\end{thebibliography}

\end{document}